# Evidence of the stability of Mo$_2$TiAlC$_2$ from first principles calculations and its thermodynamical and optical properties

Gao Qing-He[a,b], Xu Zhi-Jun[c], Tang Ling[c], Zuo Xianjun[c], Jia Guozhu[d], Du An[a*], Linghu Rong-Feng[e†], Guo Yun-Dong[f‡], Yang Ze-Jin[c,g§]

[a](College of Science, Northeastern University, Shenyang, 110004, China)

[b](Information Engineering College, Liaoning University of Traditional Chinese Medicine, Shenyang 110847, China)

[c](School of Science, Zhejiang University of Technology, Hangzhou, 310023, China)

[d](College of Physics and Electronics Engieering, Sichuan Normal University, Chengdu 610068, China)

[e](School of Physics and Electronics Sciences, Guizhou Education University, Guiyang, 550018, China)

[f](College of Engieering and Technology, Neijiang Normal University, Neijiang, 641112, China)

[g](National Lab of Superhard Materials, Jilin University, Changchun 130012, China)

[*]gqhscu@163.com
[†]linghu@gznu.edu.cn
[‡]g308yd@126.com
[§]zejinyang@zjut.edu.cn

**Abstract:**

The elastic, thermodynamic, and optical properties of $Mo_2TiAlC_2$ are investigated by first-principles calculations. Our results indicate that the *a* axis is stiffer than *c* axis within 0~100 GPa. Elastic constants calculations predict the large stability range of $Mo_2TiAlC_2$ under pressure. Several important thermodynamic properties are discussed detailedly, including the Debye temperature, thermal expansion coefficient, and heat capacity *etc*. The bonding properties are studied from the elastic quantities and electronic properties. The electronic properties are investigated, including the energy band structure, density of states, and so on. The evidence of the instability of $Mo_3AlC_2$ and stability of $Mo_2TiAlC_2$ are successfully obtained.

# 1. Introduction

The $M_{n+1}AX_n$ (n =1, 2, 3, etc) phases are a family of hexagonal compound, where M is a transition metal, A is an A–group element, and X is usually C or N. Since after Nowotny et al. synthesized a large number of MAX phases with n=1, which were called Hagg phases ( denoted as "H-phases") [1], many experimental and theoretical studies devoted to MAX phases and their solid solutions[2]. Studies shown that MAX phases combine some of the best properties of metals and ceramics[3]. This considerably rare property combination arises from the more metallic M–A, more covalent M–X bonds and the presence of mobile basal plane dislocations.

Pressure can rapidly decrease the atomic distance and thus induces the novel structural transition, non-metals becoming metals and even superconductors[4]. Under high pressure, the light alkali elements Li[5] and Na[6] experience unusual metal-to-semiconductor transition. Usually, some end members of MAX phases become thermodynamically stable by forming quaternary compounds despite that one or both of the end members is thermodynamically unstable, such as $(Cr_{2/3}Ti_{1/3})_3AlC_2$[7], $(Ti_{0.5}Nb_{0.5})_5AlC_4$[8], $(Cr_{2/3}Ti_{1/3})_3AlC_2$[9] and $(Cr_{3/8}Ti_{5/8})_4AlC_3$[9], $(Ti_{1-x}Nb_x)_2AlC$ ($x = 0$, 0.25, 0.5, 0.75, 1.0)[10]. Phase stability of $(Cr_{1-x}, M_x)_2(Al_{1-y}, A_y)(C_{1-z}, X_z)$ (M = Ti, Hf, Zr; A = Si, X=B, prototype $Cr_2AlC$) was studied using *ab initio* calculations[11], in which only $(Cr_{1-x},Ti_x)_2AlC$ compounds are stable, with *x* varying from 0 to 1 at an interval of 0.2.

Recently, $Cr_2TiAlC_2$[7, 12], $V_2CrAlC_2$[7, 12] and $Mo_2TiAlC_2$ phase[13] were synthesized. The geometry of $Mo_2TiAlC_2$ is isostructural with the other $M_{n+1}AX_n$ compound (such as $Ti_3SiC_2$) under the framework of same space group(*P6_3/mmc*)[13]. Experimental and theoretical characterizations of ordered MAX phases $Mo_2TiAlC_2$ and $Mo_2Ti_2AlC_3$[14] are reported, in which many mechanical quantities and density of states (DOS) of $Mo_2TiAlC_2$ are studied. Many Ti-doping stabilized compounds are studied through formation enthalpy[15] calculations, and their DOS calculations,which demonstrate that Al/C atomic DOSs of $Mo_2TiAlC_2$ shifting towards lower-energy side contribute to the structural stability.

Many MAX compounds present very interesting behaviors under ambient conditions or high pressure, particularly for those Ti-related ones, such as the availability of $Cr_2TiAlC_2$ and $Mo_2TiAlC_2$ through doping one Ti atom from the unstable Mo-containing $M_3AX_2$ phases. Therefore, the new synthesized $Mo_2TiAlC_2$ might have interesting technological applications. Considering its special geometric structure, it is necessary to further study its thermodynamical and optical properties under pressure.We note that the Mo-containing MAX phases known to date are $Mo_2GaC$, $Mo_2Ga_2C$ and magnetic $(Mo_{0.5}Mn_{0.5})_2GaC$. Prior to the dopant of Ti atom, Ti atom stabilized $Mo_2GaC$ presents unexpectedly c-axis ultraincompressibilities above about 15 GPa[16] as well as the hardest *c*-axis and softest *a*-axis and lowest transition pressure of $Mo_2Ga_2C$[17]. Moreover, the group member of Ti atom (such as Zr atom) formed $Zr_2InC$[18] also presents the similar *c*-axis ultraincompressibilities above about 70 GPa. These phenomena indicate that the Ti atom and its group members formed novel chemical bonds at certain conditions. Furthermore, in order to fabricate Mo-based MXenes( a new family of two-dimensional (2D) materials labeled

MXenes by selectively etching the Al from Al-containing MAX phases), further investigations on the crystal structure of $Mo_2TiAlC_2$ are important to confirm the order of atoms in M sites for MAX compounds and to lay the foundation of Mo-based MXenes.

The motivations are to study the elastic stability and electronic properties of $Mo_2TiAlC_2$ under high pressure and to search if it displays any unusual behavior under extreme conditions. First principles are reliable method in calculating the elastic and electronics properties for the condensed matters[4-6, 19-23].

## 2. Computational methods

The structural optimization are performed by the Vanderbilt-type[24] ultrasoft pseudopotential with a generalized gradient approximation[25] for the Perdew–Burke–Ernzerhof exchange-correlation function. A plane-wave with 350.0 eV energy cut-off was applied, and the 10×10×2 Monkhorst-Pack mesh was used at *k*-point sampling[26, 27]. Pseudo atomic calculations was performed for Mo $4d^5 5s^1$, Ti $3s^2 3p^6 3d^2 4s^2$, Al $3s^2 3p^1$, C $2s^2 2p^2$. The self-consistent convergence of the energy is at $5.0\times10^{-7}$ eV/atom. Since the spin-polarized option is considered during the calculations on $Mo_2GaC$[28] we therefore also used this option for $Mo_2TiAlC_2$ throughout. All the calculations are done by CASTEP[29, 30].

## 3. Results and discussion

### 3.1 Elastic properties

$Mo_2TiAlC_2$ belongs to *P6₃/mmc* and there are 12 atoms in one unit cell, the individual coordinate is: (2/3, 1/3, *u*) of Mo, (0, 0, 0) of Ti, (0, 0, 1/4) of Al, (1/3, 2/3, $v$) of C. The stable geometry parameters are obtained using the usual optimization procedure of hexagonal cell: $a=b=3.0157$ Å, $c=18.6032$ Å, $u=0.1352$, $v=0.0677$, consistent with the experimental data[13], $a=b=2.997$ Å, $c=18.6608$ Å, $u=0.13336$, $v=0.0687$, as is also shown in Table **1**. Note that the properties of $Mo_3AlC_2$ comprising of its alpha and beta phases are calculated for reference only because they are unsynthesizable experimentally at ambient conditions. All of these calculated results agree excellent with the other calculations[14] and experimental measurements[13], confirming the reliability of the present calculations. The atomic space arrangements are shown in Figure **1**.

Usually, the shrinkage along the *c* axis with pressure is faster than that along the *a* axis. Our calculations observe that *a* axis is always stiffer than *c* axis within 0~100 GPa, but the difference is vary small, as is shown in Figure **2**. The volumetric shrinkage is far faster than that of axial shrinkage.

Elastic constants determine the response of the crystal to applied forces. Our calculated elastic constants are listed in Table **1** and Figure **3**, together with other calculations[14]. Previous calculations[31] on $Cr_2GeC$, $(Cr_{0.5}V_{0.5})_2GeC$, and $V_2GeC$ failed to find the higher moduli of intermetallics in comparison with its two end members, which is different with the experimental measurement. However, the present $Mo_2TiAlC_2$ occurs abnormal phenomenon, that is, almost all of the mechanical quantities are larger than its end members. Therefore, all of the calculated elastic moduli of $Mo_2TiAlC_2$ should be confirmed by the experimental confirmation. According to the stability criteria[32], *i.e.*

$$C_{12} > 0, \ C_{33} > 0, \ (C_{11} - C_{12})/2 > 0, \ C_{44} > 0,$$
$$(C_{11} + C_{12})C_{33} - 2C_{13}^2 > 0 \tag{1}$$

It is found that $Mo_2TiAlC_2$ is mechanical stable up to about 100 GPa. The small $c_{33}$ manifests that the *c*-axis direction is relatively soft, which is consistent with the axial compressibility. As can be seen from Figure **3**, the current $c_{11} > c_{33}$ means that the atomic bonds along the [100] planes between nearest neighbors are slightly stronger than those along the [001] plane. Besides, the comparable $c_{33}/c_{11}$ and $c_{13}/c_{12}$ means that the atomic bonding along the *c* axis is similar with that of *a* axis. Meanwhile, the $c_{12}$ is always nearly equal with that of $c_{13}$ at any pressures, and $c_{44}$ presents similar variation trend with that of shear modulus *G*. These variation trends imply that the $Mo_2TiAlC_2$ should be stable in a wide pressure range.

Anisotropy factor $A = c_{33}/c_{11}$, $A=1$, meaning isotropic crystal, while any value larger or smaller than 1 indicates an elastic anisotropy. Analysis from the calculated elastic constants it is easily seen that $Mo_2TiAlC_2$ could be approximately viewed as isotropic material in a wide pressure range. The measurement of the elastic anisotropy in shear is given by the quantity $A_1 = c_{44}/c_s = 2c_{44}/(c_{11}-c_{12})$, where $c_s$ is the shear modulus corresponding to an elastic wave propagation in the [110] direction. When $c_{44}/c_s$ is equal to unity, a crystal is perfectly isotropic. The present $Mo_2TiAlC_2$ also could be approximately considered as isotropic material at 0 GPa, whereas the degree of isotropy decreases with the increasing of pressure, corresponding to 1.23 of 0 GPa and 1.91 of 100 GPa, respectively. The present result agrees with the calculated anisotropic factor, with a value of 1.27 at 0 GPa[14]. Another shear anisotropy quantity is $A_2=(c_{11}+c_{33}-2c_{13})/4c_{44}$, when $A_2=1$, a crystal is perfectly isotropic. The present variation range of $A_2$ in $Mo_2TiAlC_2$ is 0.82 of 0 GPa and 0.5 of 100 GPa, consisting well with the conclusions of $A_1$.

For covalent and ionic materials, the relations between bulk ($B$) and shear ($G$) modulus are $G\approx1.1B$ and $G\approx0.6B$, respectively. For $Mo_2TiAlC_2$ the calculated ratio of $G/B$ (using their respective average values of Voigt, Reuss and Hill estimations) are in the interval from 0.715 at 0 GPa to 0.403 at 100 GPa, indicating that the ionic bonding is more suitable for $Mo_2TiAlC_2$ at ambient conditions and the degree of ionicity increases with pressure. To evaluate material ductility or brittleness, Pugh *et al*. introduced the $B/G$ ratio[33]: the material is brittle if the ratio is less than the critical value (1.75) or the material is ductile if the ratio $G/B$ is below 0.57. Our calculated $Mo_2TiAlC_2$ are brittle ($G/B = 0.715$) at zero pressure, and the degree of brittleness decreases at high pressure. Moreover, the current $B/C_{44}$ is 1.44, within the range of $M_{n+1}AX_n$ phases 1.2-1.7. Poisson's ratio is generally used to quantify the stability of the crystal against shear and provides more information about the characteristics of the bonding forces than elastic constant. The value of Poisson's ratio for covalent materials is small (0.1), whereas for ionic materials a typical value is 0.25, and for metallic materials, a typical value is 0.33. The present Poisson's ratio is 0.2667 of 0 GPa and 0.3423 of 100 GPa, consisting with previous data, with a value of 0.25[14]. The present results means that the $Mo_2TiAlC_2$ is ionic materials and a higher metallic and ionic (or weaker covalent) contribution in inter-atomic bonding should be assumed, and the metallic or ionic increases gradually with pressure.

## 3.2 Thermodynamical properties

The thermodynamic properties of $Mo_2TiAlC_2$ are studied by the quasi-harmonic Debye model, the calculation details[34] could be obtained elsewhere. Many MAX phases will decompose at temperature above 2000 K, such as the stability limit of of $Ti_3SiC_2$[35] is 2300 K. The order-disorder transition temperature of $M_2TiAlC_2$ (M=Cr, Mo, W) is about 1773 K[15], which is far larger than the other MAX phases, meaning the important role of Ti-doping. Therefore, we calculate all these properties to about 1500 K. The Debye temperature $\Theta$ is a fundamental parameter of a material. Our calculated $\Theta$ is 708.58 K at $T$=0 K and $P$=0 GPa. From Figure **4**, we can obtain: (a) When the pressure keeps constant (0, 30, 60, 90 GPa), the $\Theta$ decreases linearly with temperature. Moreover, the Debye temperature increases with pressure increasing, and the increase magnitude descends with same pressure interval. In particular, the $\Theta$ decreases extremely slow with temperature at higher pressure. (b) When the temperature keeps constant (0, 300, 600, 1000, 1500 K), the $\Theta$ almost increases linearly with pressure. The $\Theta$ at 1500 K is slightly lower than that at 0 K and the discrepancy gradually disappears at higher pressure. These data demonstrates its strong pressure and weak temperature dependences. This similar variation trends could also be found in the Grüneisen parameter, as is shown in Figure **5**.

The variation of the thermal expansion coefficient $\alpha$ with temperature and pressure of $Mo_2TiAlC_2$ is presented in Figure **6**. In Figure **6** (a), as pressure increases, $\alpha$ decreases rapidly and the pressure dependence decreased. Moreover, the discrepancy at 1000 K and 1500 K is small and the high temperature response is rapidly decreased. In Figure **6** (b), $\alpha$ increases rapidly at low temperature and the increasing trend becomes gentler. The effects of pressure on $\alpha$ are smaller at low temperature than that at high temperature. Generally, $\alpha$ responds sensitively at lower temperature (below 500 K) and pressure (below 50 GPa), and the lower the temperature/pressure is, the faster the $\alpha$ changes. It can be found that the $\alpha$ converges to a constant value at high temperature and pressure.

The isothermal ($B_T$) and adiabatic ($B_S$) bulk moduli exist small differences owing to the very small thermal expansion coefficient $\alpha$ and Grüneisen parameter $\gamma$, $B_S=B_T(1+\alpha\gamma T)$. In Figure **7**, both $B_T$ and $B_S$ bulk moduli decrease monotonically with temperature at zero or high pressure, but $B_T$ varies significantly larger than that of $B_S$. Similarly, both $B_T$ and $B_S$ increase with pressure at zero or high temperature. The current investigations confirmed that $B_S$ is always larger than $B_T$ over a large pressure and temperature range and both of them increase with pressure at zero temperature.

According to the Debye temperature, we could calculate the lattice entropy $S$ at different pressures and temperatures, as is shown in Figure **8**. Clearly, when the pressure keeps constant, the $S$ increases monotonously with temperature. On the contrary, when the temperature keeps constant, the $S$ decreases monotonously with pressure, indicating its strong pressure and temperature dependences.

In Figure **9**, the heat capacity at constant pressure $C_P$ and at constant volume $C_V$ with temperature $T$ and pressure $P$ are calculated. The computational results indicate that $C_V$ and $C_P$ increase with increasing temperature, the specific heat obeys the Debye $T^3$ power-law behavior in the low temperature limit, and $C_V$ approaches the classical asymptotic limit of $C_V = 3nNk_B$=149.54 J/mol*K for quaternary Mo$_2$TiAlC$_2$ at the temperature above 800 K. These characteristics of heat capacity show the fact that the interactions between ions in the nanolaminates have great effect on heat capacities especially at low temperatures. Furthermore, there is a difference between $C_P$ and $C_V$ in the normal state for the phases due to the thermal expansion caused by anharmonicity effects. The difference between $C_P$ and $C_V$ is small at low temperature. At high temperature, the $C_V$ approaches a constant value, $C_P$ increases monotonously with temperature increasing. In Figure **9 (a)**, **(b)**, the discrepancy between $C_P$ and $C_V$ is negligible until 100 GPa under lower temperature (such as below 600 K), but the discrepancy slightly increased at higher temperature, such as 1500 K. Figure **9 (c)**, **(d)** reveals the minor difference of $C_P$ and $C_V$ at high temperature under 0 GPa. Under high pressure, the temperature dependences of $C_P$ and $C_V$ could be ignored reasonably due to the weak thermodynamical properties. These thermodynamical quantities present similar variation trends with the other MAX phases[36-40].

### 3.3 Electronic properties

For understanding the chemical bonding of Mo$_2$TiAlC$_2$, we have calculated the band structure and density of states (DOS) at 0 GPa and compared them with the available computations[14, 15]. As is seen from Figure **10**, together with the hypothetical Mo$_3$AlC$_2$, the total DOS of Mo$_2$TiAlC$_2$ reveals its metallic feature. The values of Mo$_2$TiAlC$_2$ and Mo$_3$AlC$_2$ at Fermi level are 5.5 and 9.9 electrons/eV, respectively. The high value of $Mo_3AlC_2$ (alpha phase) corresponds to the occurrence of phonon imaginary frequency, as is shown in Figure **11**.

The partial states (*s*, *p*, *d*) of Mo atom in $Mo_2TiAlC_2$ present quite similar distributions with those of Mo atom located along the *c* axis in $Mo_3AlC_2$, only the *d* states display small discrepancy in the vicinity of Fermi level, as is shown in Figure **10** (**b**), (**c**). Such small discrepancy is caused by the different chemical environments as the Mo atom still occupies the same positions in the two compounds, namely, only those Mo atoms sited in the *ab* plane are substituted, as is also shown in Figure **1**. In Figure **10** (**d**), the partial states (*s*, *p*, *d*) of Mo atom locating at *ab* plane in $Mo_3AlC_2$ show evident difference with those of Mo atom locating along *c* axis in $Mo_2TiAlC_2$. These differences include the relative shifts of *s* and *p* states and a profile change of *d* state. In Figure **10** (**b**), (**d**), a relative shift of *s* and *p* states of Mo atom towards higher-energy side in $Mo_2TiAlC_2$ occurs in comparison with the correspondent states of Mo atom locating at *ab* plane in $Mo_3AlC_2$. However, such variation is far smaller than that of Ti atom, as is shown in Figure **10** (**d**), (**e**). Moreover, the small shift between the different Mo atoms for their same positions along *c* axis could be ignored reasonably, as is shown in Figure **10** (**b**), (**c**). This small shift denotes that these states are irrelativity to the structural stabilization. Unexpectedly, the value of *d* state of Ti atom at Fermi level is about 0.23 electrons/eV, which is far smaller than that of Mo atom at the same occupations, with a value of 1.25 electrons/eV shown in Figure **10** (**d**) (**e**). Such significant decrease, which is far larger than previous observations[15], can effectively stabilize the bond strength. This substantial decrease was partially attributed to the less *d* orbital electrons of Ti atom. The *s* and *p* states of C atoms present a global shift towards lower energy side in $Mo_3AlC_2$ in comparison with those of $Mo_2TiAlC_2$, playing secondary role in stabling the lattice of $Mo_3AlC_2$. In fact, the global shift of *s* and *p* states of C atoms in $Mo_2TiAlC_2$ might cause more charge overlap with *d* states of Ti and Mo atoms and therefore form stronger covalent bond. However, as is shown in Figure 10 (**f**), the negligible shift of *s* and *p* states of Al atoms between $Mo_2TiAlC_2$ and $Mo_3AlC_2$ possibly relates to its special in-plane occupancies, whereas previous calculations[15] observed the lower energy sites of the bonding states between Mo *d* and Al *p* in $Mo_2TiAlC_2$. The Mo *d*, Ti *d*, Al *p*, and C *p* electrons co-contribute to the energy range of -10~0 eV.

Energy band analysis observes that four (51-54) and eight (51-58) orbitals cross Fermi level in $Mo_2TiAlC_2$ and $Mo_3AlC_2$, respectively, completely consisting with the fact of high DOS value of $Mo_3AlC_2$ at Fermi level, as is shown in Figure **12** (**a**) (**b**). The large shift (about 0.75 eV) of *G* point towards high-energy side in $Mo_2TiAlC_2$ will strongly repulse the electrons shift to low-energy zone, making more electrons of Ti *d* states accumulation to the Ti-C bonding orientation and the edge of rhombic *ab* plane, totally consistent with the analysis of DOS in Figure **10** (**d**), (**e**). Moreover, less *d* orbital electrons of Ti atom built simpler energy level distributions around Fermi level, which could reduce the number of anti-bonding state. In particular, such unusual energy shift of *G* point will lead to stronger Ti-C bond due to the local energetic discrepancy, as is also the case of the energy distribution feature of *A* point, both of which are far lower than those of *H* and *K* points in energy, as is shown in Figure **12**. The coordinates of the special points are *G* (0, 0, 0), *A* (0, 0, 0.5), *H* (-0.333, 0.667, 0.5), *K* (-0.333, 0.667, 0), *M* (0, 0.5, 0), and *L* (0, 0.5, 0.5), respectively. This phenomenon is consistent with the facts of the orbital analysis, as is shown in Figure **13** (**a**) (**b**). Orbitals 51-54 demonstrate that little Ti *d* states of $Mo_2TiAlC_2$ contribute to the bonding, totally different with the cases of $Mo_3AlC_2$. For better understanding the electron charge transfer direction, we also calculate the electron density differences and electron localization functions for the two compounds, as is shown in Figure **14**. It is found that the substitution of Mo by Ti atom has led to significant charge transfer between C, Ti, and Mo atoms and formed stronger bond in $Mo_2TiAlC_2$.

**3.3 Optical properties**

To clarify the structural related properties, we also calculate the optical properties for radiation energy up to 45 eV via the absorption, reflectivity, energy loss, refractivity, dielectric, and conductivity coefficients *etc* at 0 GPa. The linear optical properties in solids can be described by the complex frequency-dependent dielectric function $\varepsilon(\omega) = \varepsilon_1(\omega) + i\varepsilon_2(\omega)$.

The imaginary part $\varepsilon_2(\omega)$ of the frequency-dependent dielectric function is given by

$$\varepsilon_2(\omega) = \frac{e^2\hbar}{\pi m^2 \omega^2} \sum_{v,c} \int_{BZ} |M_{vk}(k)|^2 \delta[\omega_{cv}(k) - \omega] d^3k. \qquad (1)$$

The integral is performed over the first Brillouin zone, and the momentum dipole elements $M_{vk}(k) = \langle U_{cv} | e \cdot \nabla | U_{ck} \rangle$, where *e* is the potential vector defining the electric field, are matrix elements for direct transitions between valence-band $u_{vk}(r)$

and conduction band $u_{ck}(r)$ states, and the $h\omega_{cv}(r) = E_{ck} - E_{vk}$ is the corresponding transition energy.

The real part $\varepsilon_1(\omega)$ could be derived from the imaginary part using the Kramers-Kronig relations,

$$\varepsilon_1(\omega) = 1 + \frac{2}{\pi} p \int_0^\infty \frac{\omega' \varepsilon_2(\omega')}{\omega'^2 - \omega^2} d\omega', \qquad (2)$$

where $p$ denotes the Cauchy principal value of the integral. Using the knowledge of real and imaginary parts of the frequency-dependent dielectric function, we thus can calculate the other optical functions.

The current incident photon polarization directions are (1, 0, 0) and (0, 0, 1), named $E \perp c$ and $E//c$ in these figures, respectively. The optical properties can be conveniently divided into two spectral regions, the low-energy range of 0~15 eV and the high-energy range of 35~40 eV, respectively.

In Figure **15**, the large reflectivity index at low-energy range illustrates the strong optical response, consisting with the sensitivities of other properties. The calculated linear absorption spectrum presents two prominent peaks and the absorption edge starts from about 0 eV, corresponding to the energy pseudogap. The absorption thresholds along the two polarization direction are nearly the same. The absorption region is quite wide, the absorption intensity is quite large, which would make it a potential candidate for photoelectron application. The complex refractive index is composed of the real *n* and the imaginary part *k*. In the high energy range, such as above about 35 eV, *n* almost becomes a constant and *k* also decrease to zero, which show the system is relatively weak in absorption of high frequency wave. The dielectric function presents similar variation trends with those of refractive index, its giant real part is clearly seen. The peaks at about 1 eV correspond probably to the transition between the Al *p* and C *s* states or the Ti *p* and Mo *d* states, where the ambiguous origin is attributed mainly to their similar peak intervals. Both the real and imaginary parts present onefold positive value. The calculated real and imaginary parts of the conductivity index are also presented. Both the real and imaginary parts respond sensitively in the energy range of 0~15 eV and 35~40 eV, respectively. The peaks of energy loss function correspond to the trailing edges of the refractive index *n*. The reflectivity also corresponds well to the energy loss spectrum.

## 4. Conclusion

We have performed comprehensive elastic, thermodynamic, electronic, and optical properties investigations for $Mo_2TiAlC_2$. The elastic properties confirmed the structural stability of $Mo_2TiAlC_2$ within 0~100 GPa. The structural instability of $Mo_3AlC_2$ for its two different phases are confirmed by the elastic constants and phonon dispersion curve. Furthermore, calculations on the energy band, density of states (DOS), orbitals, charge loss and gain features demonstrate that the high-energy zone of *ab* plane or rhombic center will promote the formation of stronger Ti-C bond. The fact of eight and four energy levels crossing Fermi level in $Mo_3AlC_2$ and $Mo_2TiAlC_2$ reveals the role of Ti atom in stabling the lattice during its substitution. The DOS of Ti atom determined the most likely origin of the lattice stability owing to its less *d* orbital electrons and simpler energy level distributions, as well as its special occupancies in the lattice.

**Acknowledgement**

Projects supported by the Natural Science Foundation of China (Grant No: 11304279, 11364007), Natural Science Foundation of Zhejiang Province, China (Grant Nos: LY13A040004, LY16A040013), China Postdoctoral Science Foundation (Grant No. 2012M520666), Science and Technology Foundation from Ministry of Education of Liaoning Province (Grant No: L2015333), and the Science and Technology Foundation from Guizhou Province (Grant No: J [2013]2242).

Table **1**. The structural parameters and mechanical properties of $Mo_2TiAlC_2$. Note that the Voigt, Reuss, and Hill estimations of shear modulus $G$ and Lame Lambda are different, but the other moludi including the bulk modulus $B$, Young's modulus $E$ on its three axial components (x, y, z) are identical. Thu units of these mechanical quantities are GPa, the lattice parameters are Å, and the volume is Å$^3$, respectively.

| | $C_{11}$ | $c_{33}$ | $c_{44}$ | $c_{12}$ | $c_{13}$ | $B$ | $G$ | Lame | $E_{x=y}$ | $E_z$ | $a$ | $c$ | $V$ |
|---|---|---|---|---|---|---|---|---|---|---|---|---|---|
| $Mo_2TiAlC_2$ | 384 | 371 | 150 | 140 | 130 | 216 | 133 | 127 | 312 | 306 | 2.9987 | 18.7588 | 146.09 |
| Cal.[14] | 386 | 367 | 150 | 143 | 140 | 221 | 132 | | | | 3.0082 | 18.757 | |
| Exp.[13][a] | | | | | | | | | | | 2.997 | 18.661 | |
| | | | | | | | | | | | 2.93 | 18.9 | |
| $Mo_3AlC_2$ (alpha phase) | 373 | 350 | 132 | 143 | 131 | 212 | 122 | 130 | 296 | 183 | 3.0616 | 18.6170 | 151.1276 |
| Cal.[14] | 354 | 352 | 110 | 143 | 149 | 216 | 106 | | | | 3.0714 | 18.542 | 151.48 |
| $Mo_3AlC_2$ (beta phase) | 289 | 368 | -200 | 89 | 86 | 163 | -14 (Voigt) 279 (Reuss) 132 (Hill) | 173 (Voigt) -23 (Reuss) 74 (Hill) | 251 | 329 | 3.089 | 19.8849 | 164.3323 |

[a]: the lattice parameters are measured from two different methods, the values of 2.997 and 18.757 are obtained from the X-ray diffraction patterns, and the values of 2.93 and 18.9 are obtained form Selected area electron diffraction (SAED) characterization.

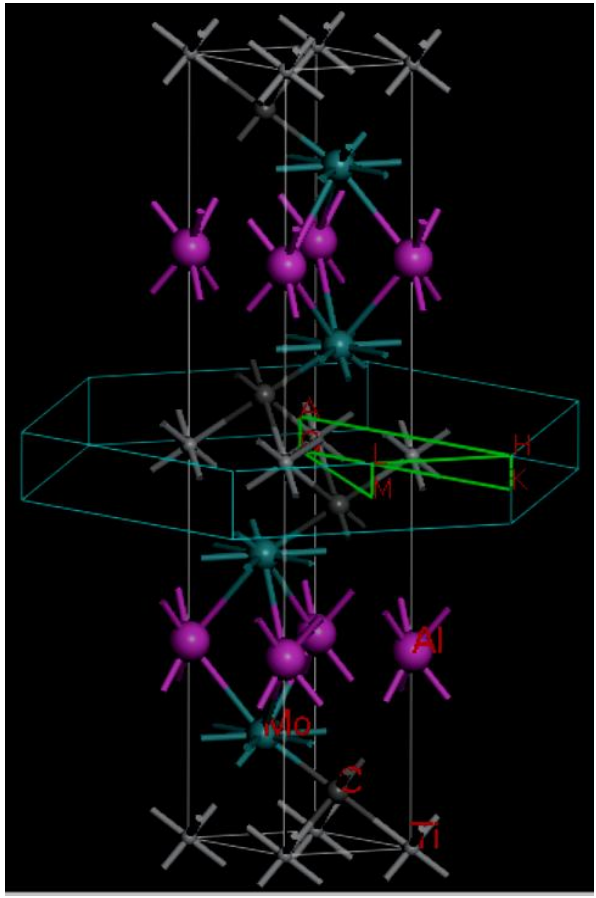

Figure 1. The lattice structure[11,13,14] of Mo$_2$TiAlC$_2$, together with the Brillouin zone, the horizontal lines are *ab* plane, and the vertical line is *c* direction. Mo$_3$AlC$_2$ can be obtained by replacing the Ti atoms in the *ab* plane by Mo atoms.

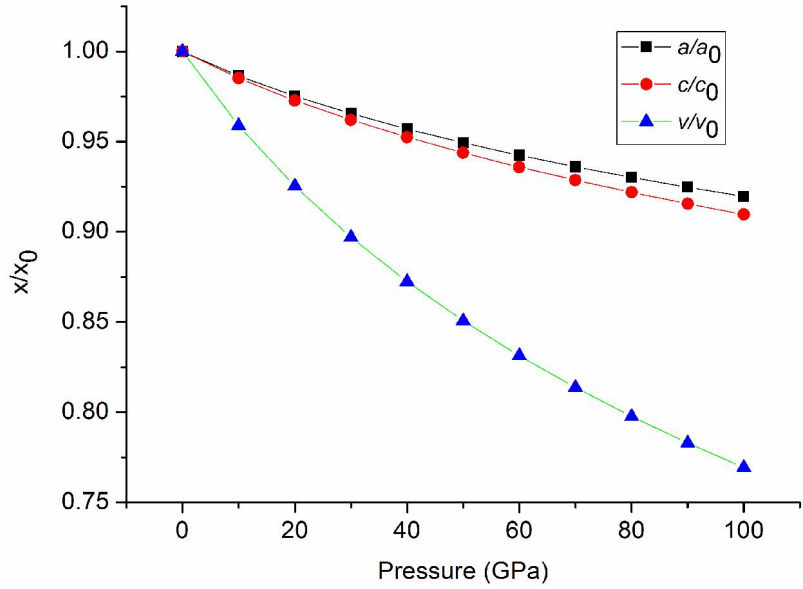

Figure 2. (Color online only). Axial compressibilities of *a*, *c*, and volumetric shrinkage of $Mo_2TiAlC_2$, where *X* represents *a*, *c*, and *v* at any pressures, $X_0$ represents *a*, *c*, and *v* at zero pressure.

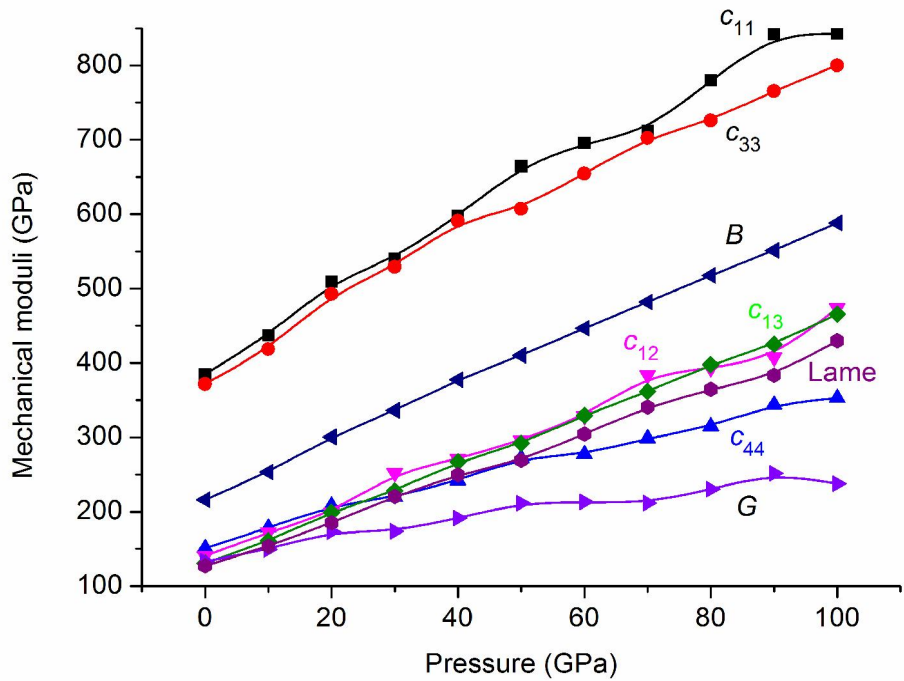

Figure 3. The elastic constants and mechanical moduli of $Mo_2TiAlC_2$ within 0~100 GPa at 0 K.

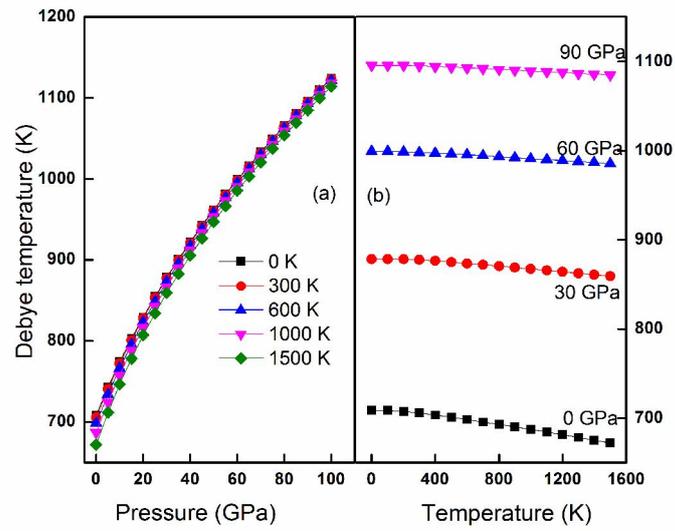

Figure 4. Debye temperature as a function of Mo$_2$TiAlC$_2$ under different temperatures and pressures.

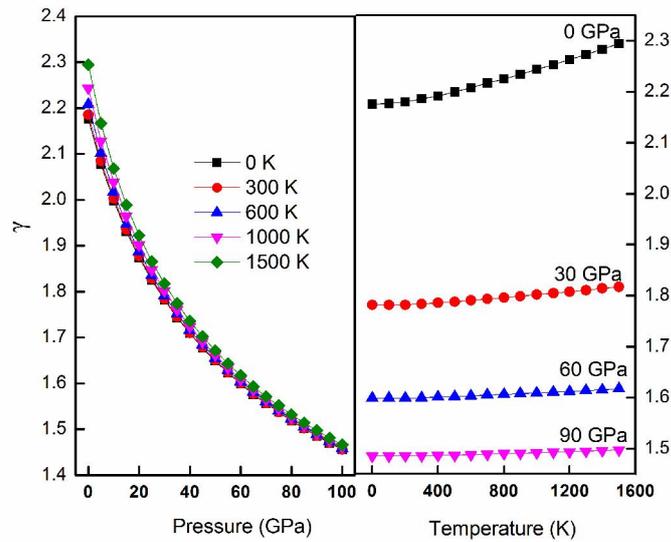

Figure 5. Grüneisen parameter as a function of Mo$_2$TiAlC$_2$ under different temperatures and pressures.

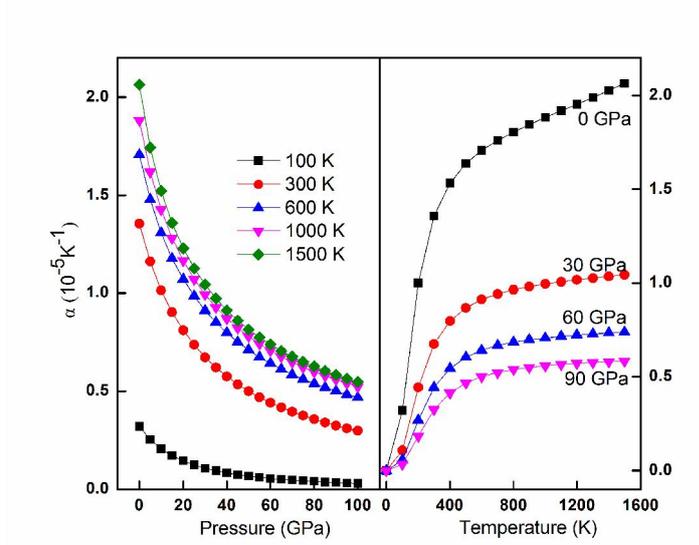

Figure 6. Thermal expansion coefficient of $Mo_2TiAlC_2$ as a function of temperature and pressure.

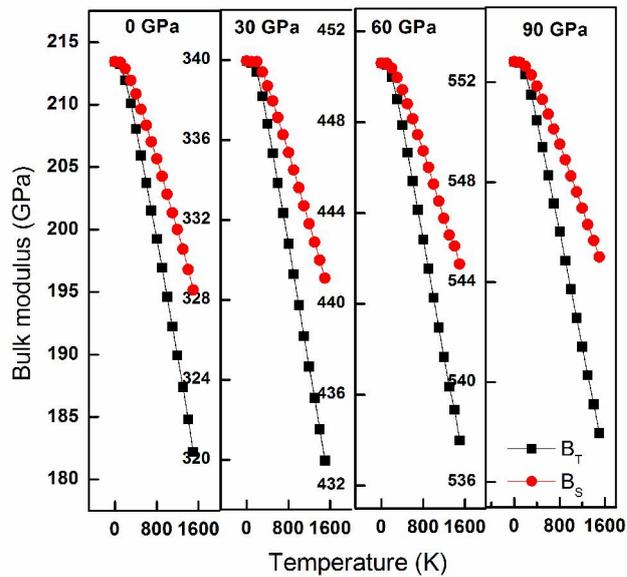

Figure 7. Isothermal ($B_T$) and adiabatic ($B_S$) bulk modulus of $Mo_2TiAlC_2$ under different temperatures and pressures.

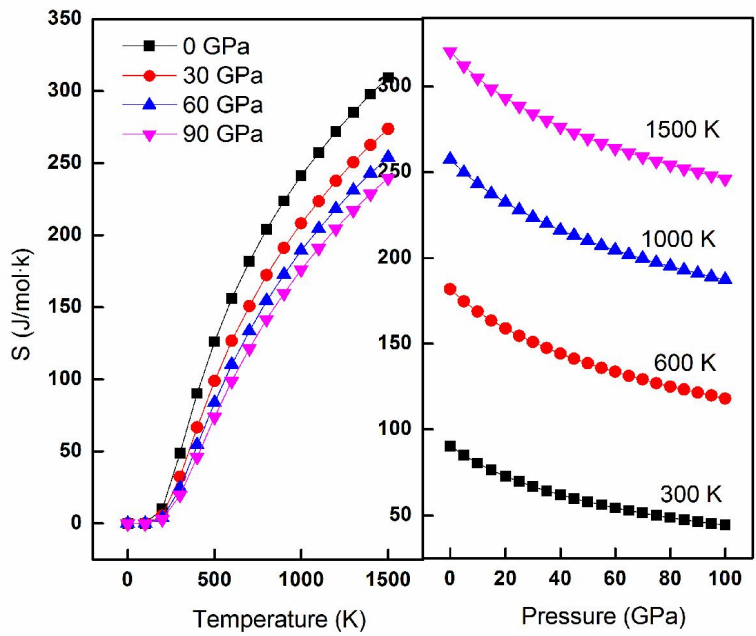

Figure 8. Lattice entropy as a function of Mo$_2$TiAlC$_2$ under different temperatures and pressures.

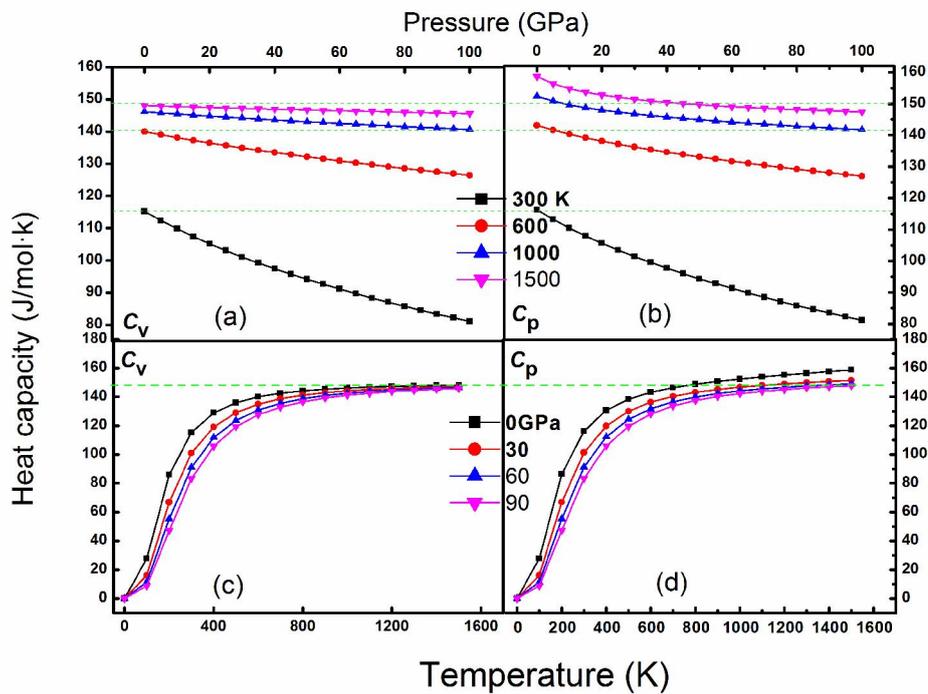

Figure 9. Heat capacity of Mo$_2$TiAlC$_2$ under different temperatures and pressures.

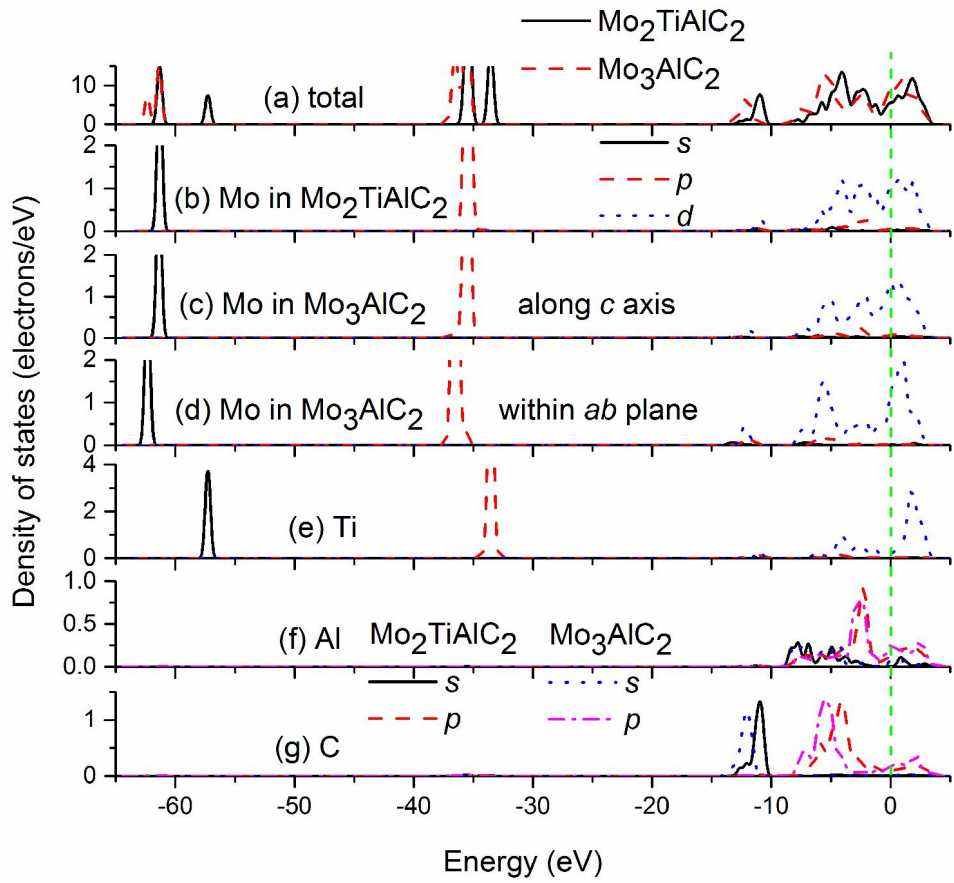

Figure 10. The density of states of $Mo_2TiAlC_2$ and the hypothetical $Mo_3AlC_2$.

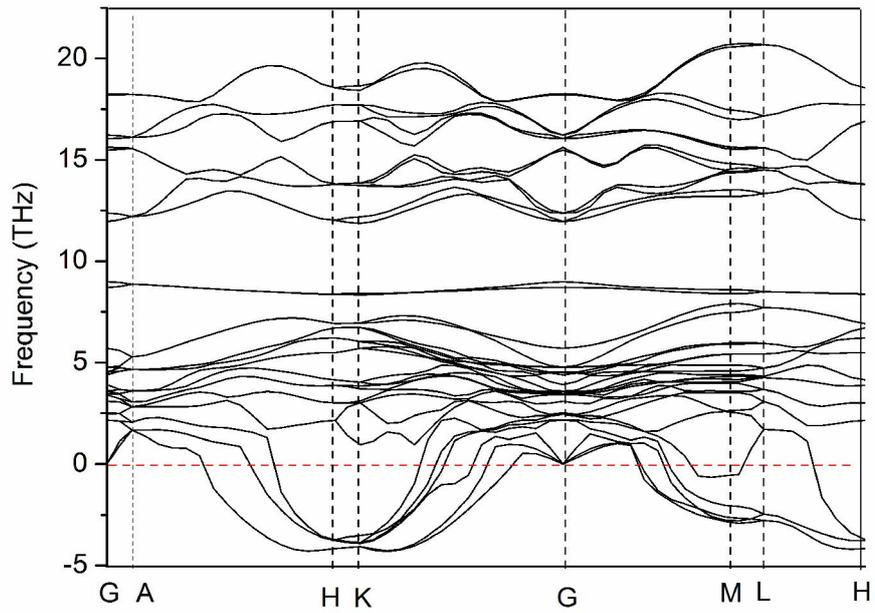

Figure 11. Phonon dispersion curve of $Mo_3AlC_2$ (alpha phase) at 0 GPa.

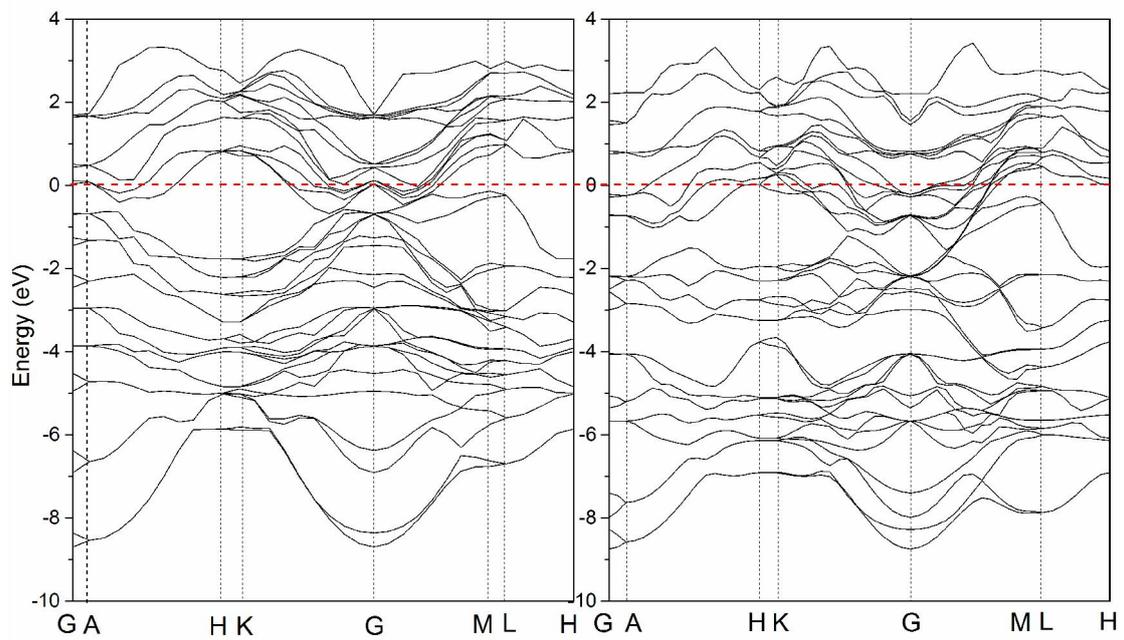

Figure 12 (a). Energy bands of $Mo_2TiAlC_2$ (left) and $Mo_3AlC_2$ near Fermi level.

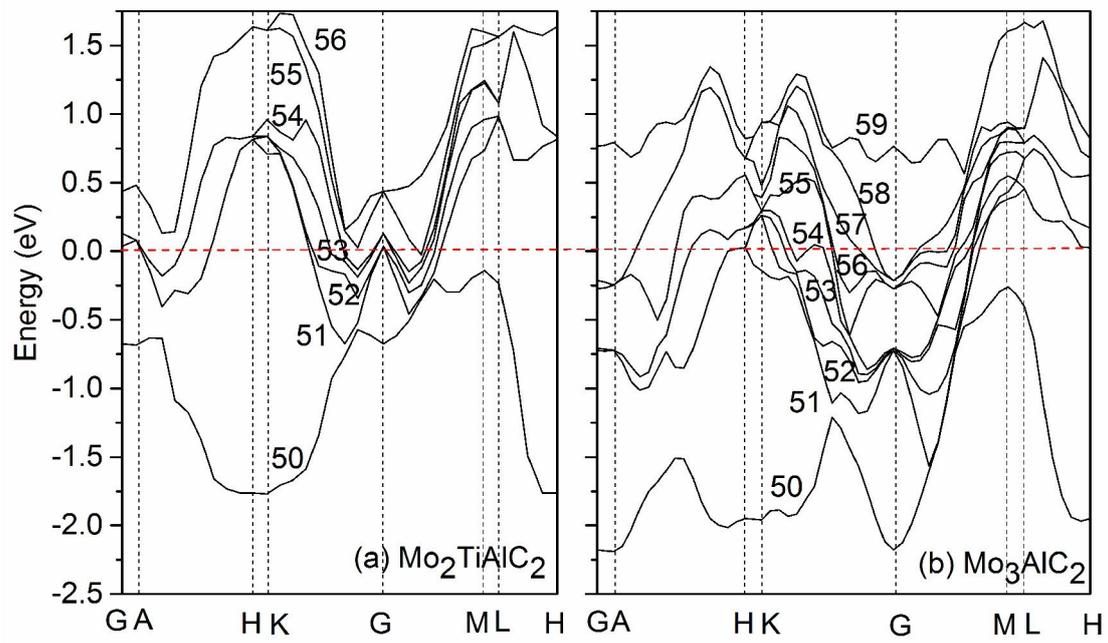

Figure 12 (b). Energy bands of Mo$_2$TiAlC$_2$ (left) and Mo$_3$AlC$_2$ that cross the Fermi level.

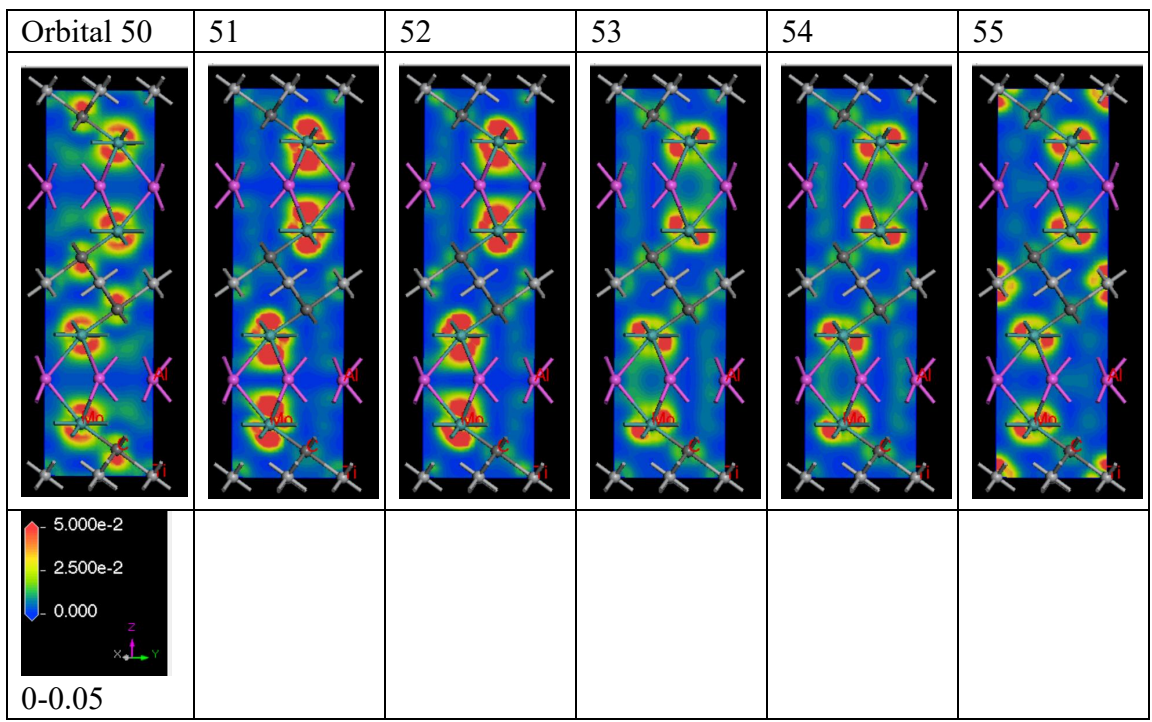

Figure 13 (a). Orbitals crossing Fermi level are shown in Mo$_2$TiAlC$_2$, together with their nearest neighbors 50 and 55. The color range is within 0~0.05 starts from blue, green, yellow, and red in turn. The horizontal axis *x* or *y* represents lattice *a* or *b* direction, vertical axis *z* means *c* direction.

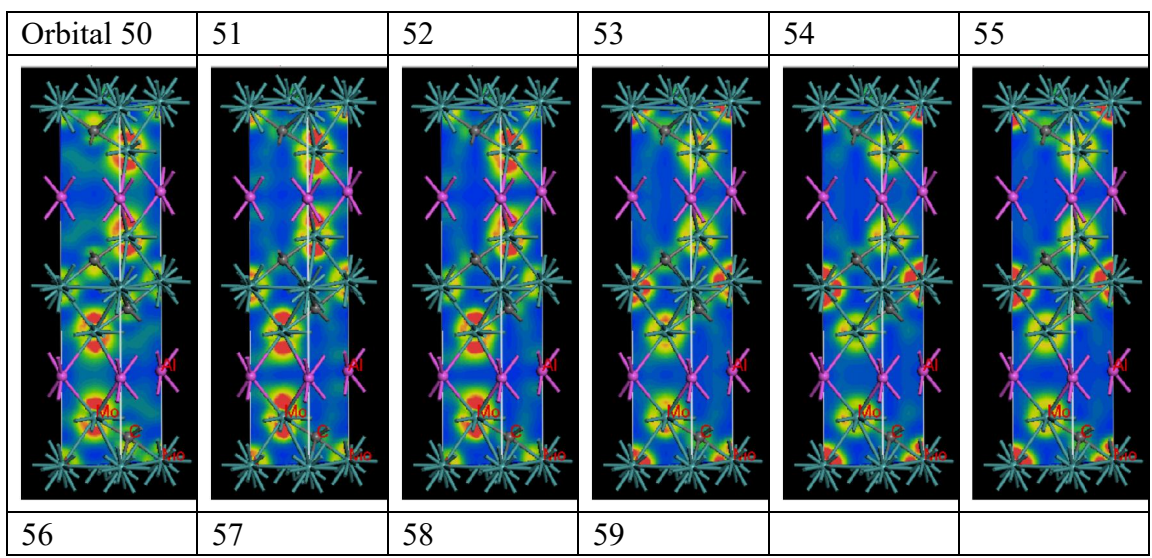

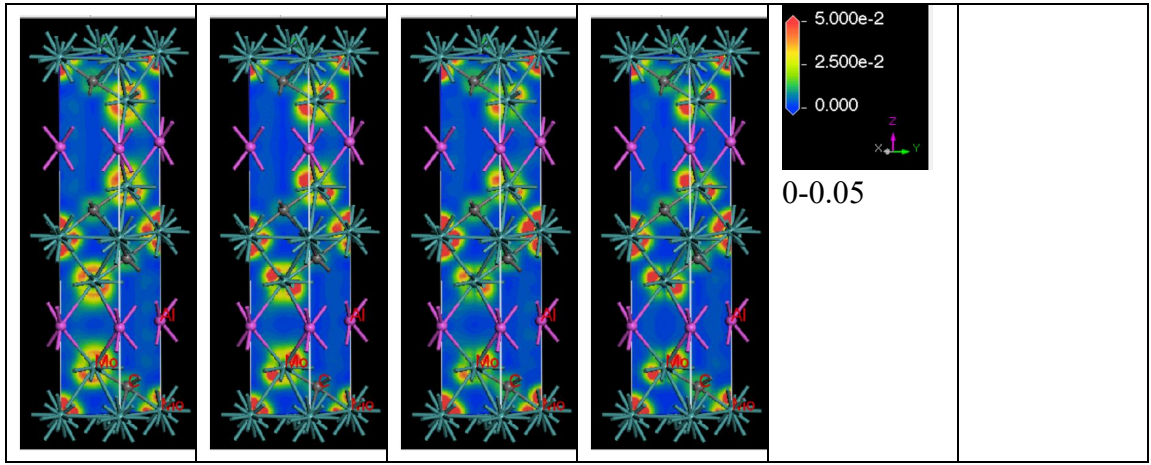

Figure 13 (b). Orbitals crossing Fermi level are shown in $Mo_3AlC_2$, together with their nearest neighbors 50 and 59. The color range is same with that of Figure 13 (a).

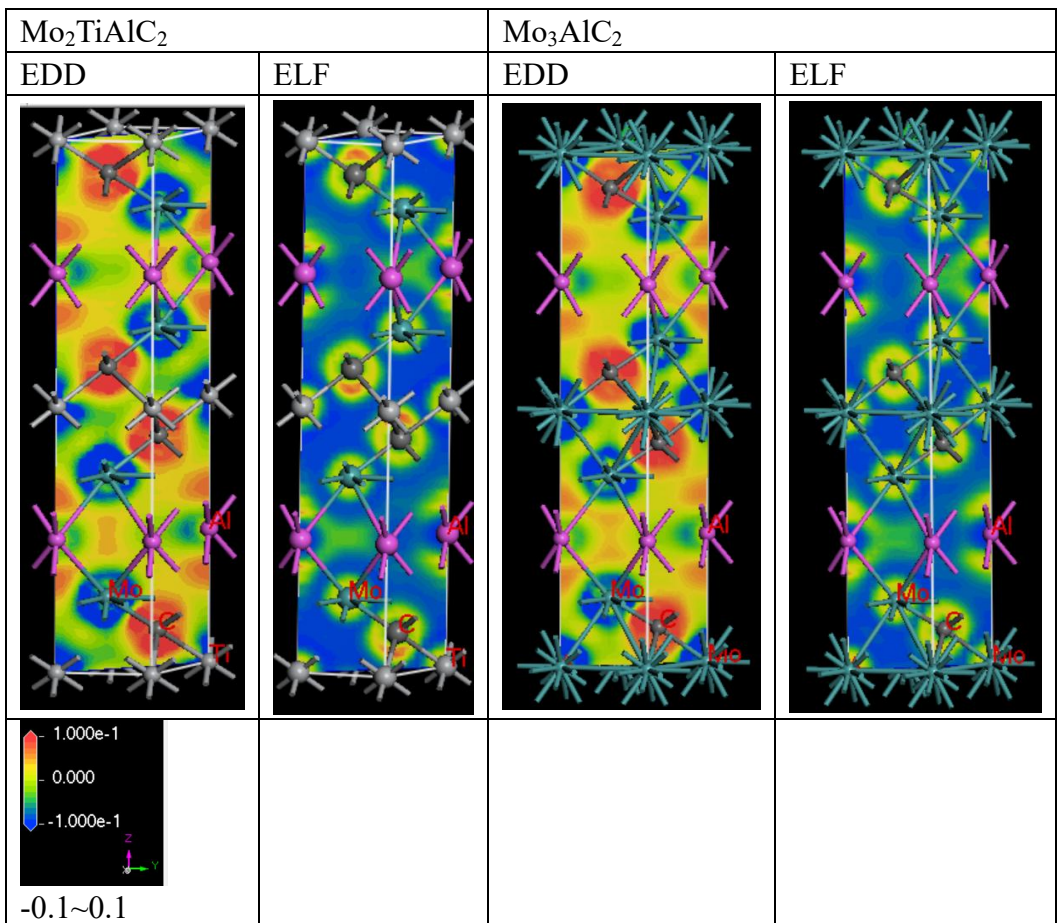

Figure 14. Electron density differences (EDD) and electron localization functions (ELF) of $Mo_2TiAlC_2$ and $Mo_3AlC_2$. The color range is same with that of Figure 13 with the only exception of larger digital range -0.1~0.1.

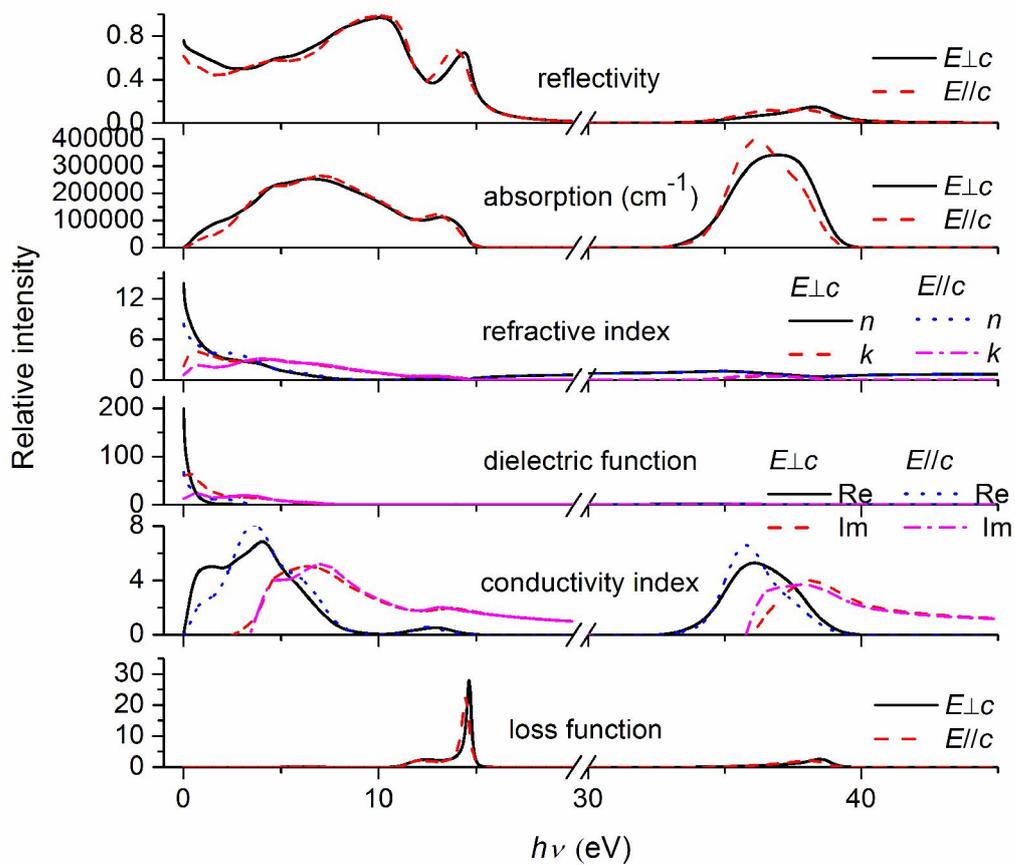

Figure 15. Optical properties of $Mo_2TiAlC_2$.